\documentstyle[osa]{revtex} 

\begin{document}
\title{Does long-range antiferromagnetism help or inhibit superconductivity?}
\author{Liliana Arrachea$^{a, b}$ and A.A. Aligia$^{c}$.}
\address{$^{a}$ Max-Planck Institut f\"{u}r Physik komplexer Systeme,\\
N\"othnitzer Str. 38, 001187 Dresden, Germany\\
$^{b}$ Institut f\"ur Theoretische Physik der Freien Universit\"at Berlin,\\
Arnimallee 14, 14195 Berlin, Germany\\
$^{c}$ Centro At\'{o}mico Bariloche and Instituto Balseiro,\\
Comisi\'on Nacional de Energ\'{\i}a At\'{o}mica, \\
8400 Bariloche, Argentina.}
\maketitle

\begin{abstract}
We analyze the possible existence of a superconducting state in a background
with long-range antiferromagnetism. We consider a generalized Hubbard model
with nearest-neighbor correlated hopping in a square lattice. Near
half filling, the model exhibits a $d$-wave-Bardeen-Cooper-Schrieffer (BCS)
solution in the paramagnetic state. 
The 
superconducting solution would be enhanced by the antiferromagnetic background
if the contribution of triplet
pairs with $d$-wave symmetry and total momentum $(\pi ,\pi )$
could be  neglected. 
However, we find that due to their contribution, the coexistence of
superconductivity and long-range antiferromagnetism is ruled out for large
values of the Coulomb repulsion $U$. Spin-density wave fluctuations (SDWF)
do not change this result.
\end{abstract}


\twocolumn

\section{Introduction.}

Eleven years after the discovery of the high-$T_{c}$ superconductivity \cite
{be-mu} several questions concerning the pairing mechanism and other
important features of the phase-diagram of the superconducting cuprates
remain without a satisfactory answer. Fortunately, there are also some
aspects of the complex nature of these materials that became quite clear. It
is widely accepted that the antiferromagnetic correlations play an important
role in the physics of the superconducting phase and there is a good amount
of evidences in favor of the $d$-wave symmetry of the superconducting order
parameter.

The proximity between the superconducting and antiferromagnetic phases in
the phase diagram of the cuprates inspired the theoreticians of this field
in different ways. In some theories an antiferromagnetic background with
long-range order is assumed to provide the scenario for the pairing \cite
{in,schri,naz,pla}. In other class of theories \cite{an,pines,rvb}, the
pairing takes place in a background of short-range antiferromagnetic
correlations with coherence lengths of only a few lattice sites.
Furthermore, the different phases of the cuprates have been considered to be
the result of the competition between these two different kinds of order 
\cite{so5}.
 The coexistence
between antiferromagnetism and superconductivity has been 
discussed in other systems with
conventional electron-phonon interaction \cite{fi,zwi} or in heavy fermion
compounds \cite{gei}. In particular, some Chevrel phase compounds containing
rare earths, exhibit an anomaly in the dependence of the upper critical
field with temperature, as the system becomes antiferromagneticaly
 ordered \cite{ishi}. This
behavior has been explained using Eliashberg theory \cite{zwi} and is due to
the change in the quasiparticle-phonon interaction in the magnetically
ordered phase.

The model we consider is defined by the Hamiltonian 
\begin{eqnarray}
H &=&U\sum_{i}n_{i\uparrow }n_{i\downarrow }\;-\;\mu \sum_{i}(n_{i\uparrow
}+n_{i\downarrow })  \nonumber \\
&-&\sum_{<ij>\sigma }(c_{i\bar{\sigma}}^{\dagger }c_{j\bar{\sigma}%
}+h.c)\{t_{AA}\;(1-n_{i\sigma })(1-n_{j\sigma })\nonumber \\
&+&t_{BB}\;n_{i\sigma
}n_{j\sigma }  \nonumber \\
&+&t_{AB}\;[n_{i\sigma }(1-n_{j\sigma })+n_{j\sigma }(1-n_{i\sigma })]\},
\label{1}
\end{eqnarray}
where $<ij>$ denotes nearest-neighbor positions of the lattice. It was
obtained from a reduction of the three-band extended Hubbard model to
describe the low-energy physics of the superconducting cuprates \cite
{sim,sch}. The values of three different hopping integrals, $%
t_{AA},t_{AB},t_{BB}$, depend on the values of the parameters of the
three-band Hamiltonian, i.e. on the Cu-O hopping $t_{pd}$, the
charge-transfer energy $\Delta $ and the on-site Cu Coulomb repulsion $U_{d}$%
. In this paper, we consider the case $t_{AB}>$ $t_{AA}=t_{BB}$, which
corresponds to the limit $t_{pd}\ll \Delta $ and $t_{pd}\ll U_{d}-\Delta $.
This minimal one-band model was useful to study several features of the
normal state of the superconducting cuprates \cite{sim,sch,opt}. A more
accurate description requires the addition of a next-nearest neighbor
hopping \cite{rvb,fei}, which we neglect here for simplicity.

Mean-field Hartree-Fock (HF) and Bardeen-Cooper-Schrieffer (BCS)
approximations are the simplest approaches to the study of correlated
systems. Although these techniques are not expected to be valid in the
strong coupling regime ($U\rightarrow \infty $) \cite{ruck} they can give
valuable insight within the weak to intermediate coupling ones. In particular, the
transition to the insulating spin-density-wave (SDW) phase that takes place
at finite $U$ for $t_{AB}<$ $t_{AA}=t_{BB}$ is described with the SDW-HF
approximation with acceptable quantitative precision \cite{mit,fest}. We
previously studied the phase diagram of (\ref{1}) using BCS and SDW-HF for
the same relation of hopping parameters considered here \cite{physc}. We
found that the correlated-hopping term gives rise to an effective pairing
interaction with components in the $s$- and $d$-wave channels. While the $s$%
-wave BCS solution is stable for low densities, the BCS solution with $d$%
-wave symmetry exists in the paramagnetic phase near half filling, within
the same range of densities as the SDW-HF solution. The latter solutions are
nearly degenerate at $U=0$. However, for finite $U$ the SDW solution is the
one with the lowest energy.

The aim of this work is to investigate the possible coexistence of both
kinds of order near half filling. The density of states in the split band
structure of the antiferromagnetic state has a van Hove singularity at
negative energies and it is further enhanced by the
opening of an antiferromagnetic gap. This situation certainly enhances the
magnitude of the BCS-gap within a range of densities close to that
corresponding to the optimal doping of the cuprates. Such a favorable
situation is the main ingredient of the first class of theories above
mentioned. However, due to the broken symmetry of the antiferromagnetic
state, not only singlet but also triplet pairs are coupled by the attractive
interaction. We show that for certain kind of interactions, like the one
considered here, or the exchange interaction $J\sum_{<ij>}{\bf S}_{i}\cdot 
{\bf S}_{j}$ of $t-J$-like models, both contributions have opposite sign and
tend to cancel in the strong coupling limit. We also show that this tendency
is not modified by spin-density wave fluctuations (SDWF) treated within the
random-phase approximation (RPA). This kind of behavior might be related to
the physics of the underdoped materials.

In section 2 we describe the mean-field picture and discuss the results.
Section 3 contains the analysis of the RPA fluctuations. We summarize and
interpret our results, and compare them with other theories in section 4.

\section{Mean-field solution.}

It is convenient to separate the correlated hopping terms of the Hamiltonian
(\ref{1}) in one- two- and three-body contributions. The complete
Hamiltonian reads 
\begin{eqnarray}
H&=&U\sum_{i}n_{i\uparrow }n_{i\downarrow }+\sum_{<ij>\sigma }
(c_{i\bar{\sigma}%
}^{\dagger }c_{j\bar{\sigma}}+h.c.)\{-t \nonumber\\
& + &t_{2}\;(n_{i\sigma }+n_{j\sigma
})+t_{3}\;n_{i\sigma }n_{j\sigma }\},  \label{2}
\end{eqnarray}
where $t=t_{AA}$, $t_{2}=t_{AA}-t_{AB}$ and $t_{3}=2t_{AB}-t_{AA}-t_{BB}$.
Details about the mean-field decoupling of the three-body terms can be found
in \cite{mit,fest,physc,fog}. For simplicity, in what follows, we do not
take into account BCS terms with $s$-wave symmetry. As shown in Ref. \cite
{physc}, these terms are relevant only for  very high doping.

The effective one-body Hamiltonian can be written as: 
\begin{eqnarray}
H_{MF} &=&C-\sum_{i\sigma }[\mu _{ef}\sigma e^{i{\bf Q\cdot R}%
_{i}}(Um/2+4t_{3}\tau )]n_{i\sigma }  \nonumber \\
&&-t_{eff}\sum_{<ij>\sigma }(c_{i\sigma }^{\dagger }c_{j\sigma }+h.c.)
\nonumber \\
& + &
\sum_{i,({\delta }={\bf x,y})}b_{i\delta }(c_{i+{\bf \delta }\uparrow
}^{\dagger }c_{i\downarrow }^{\dagger }-c_{i+{\bf \delta }\downarrow
}^{\dagger }c_{i\uparrow }^{\dagger }+h.c.),  \label{3}
\end{eqnarray}
where 
\begin{eqnarray}
\mu _{ef} &=&\mu -(Un/2+8t_{2}\tau +4t_{3}\tau n)  \nonumber \\
t_{eff} &=&t-t_{2}n+t_{3}[3\tau ^{2}+\varphi _{ix}^{2}-(n^{2}-m^{2})/4] 
\nonumber \\
b_{i\delta } &=&-2t_{3}\tau \varphi _{i \delta}  \nonumber \\
C/L &=&-U(n^{2}-m^{2})/4-8t_{2}n\tau \nonumber\\
& + & 4t_{3}[4\tau (\varphi _{ix}^{2}+
\tau
^{2})+m^{2}-n^{2}].  \label{4}
\end{eqnarray}
The staggered magnetization $\langle n_{i\uparrow }\rangle -\langle
n_{i\downarrow }\rangle =me^{i{\bf Q\cdot R}_{i}}$, $\tau =\langle c_{i+{\bf %
\delta}\sigma }^{\dagger }c_{i\sigma }\rangle $, and $\varphi _{i\delta
}=\langle c_{i+{\bf \delta }\uparrow }^{\dagger }c_{i\downarrow }^{\dagger
}\rangle $, with ${\bf \delta }={\bf x,y}$, must be determined
self-consistently. $n=n_{\uparrow }+n_{\downarrow }$ is the particle
density, $L$ is the number of sites of the lattice, ${\bf Q}=(\pi ,\pi )$
and ${\bf R}_{i}$ denotes the atomic position. Note that $\varphi
_{ix}=-\varphi _{iy}$ for the solution with $d$-wave symmetry.

In the paramagnetic phase $m=0$ and the amplitude of the $d$-wave BCS gap $%
\varphi _{ix}$ is independent of $U$ \cite{physc}. The critical temperature $%
T_{c}$, as a function of doping is shown in Fig. 1. As within this range of
densities the SDW solution ($m\neq 0$ and $\varphi _{ix}=0$) has lower free
energy that the BCS one, we consider a mean field solution with both order
parameters different from $0$. 
Due to the symmetry of the SDW background, the values of $\varphi _{ix}$
depend on whether $i$ belongs to the sublattice in which the $\uparrow $ or $%
\downarrow $ spins are the majority ones. The Fourier transforms of the self
consistent parameters of this solution are:
\begin{eqnarray}
n &=&\frac{1}{L}\sum_{k\sigma }^{\prime }\;(\langle c_{k\sigma }^{\dagger
}c_{k\sigma }\rangle +\langle c_{k+Q\sigma }^{\dagger }c_{k+Q\sigma }\rangle
),  \nonumber \\
m &=&\frac{2}{L}\sum_{k}^{\prime }\;\langle c_{k\uparrow }^{\dagger
}c_{k+Q\uparrow }\rangle   \nonumber \\
\tau  &=&\frac{1}{L}\sum_{k}^{\prime }\;e^{i{\bf k\cdot \delta }}(\langle
c_{k\sigma }^{\dagger }c_{k\sigma }\rangle -\langle c_{k+Q\sigma }^{\dagger
}c_{k+Q\sigma }\rangle )  \nonumber \\
\varphi _{ix} &=&\frac{1}{L}\sum_{k}^{\prime }e^{ik_{x}\delta _{x}}
[(\langle
c_{k\uparrow }^{\dagger }c_{-k\downarrow }^{\dagger }\rangle -\langle
c_{k+Q\uparrow }^{\dagger }c_{-(k+Q)\downarrow }^{\dagger }\rangle )\nonumber\\
& + & e^{i
{\bf Q\cdot R}_{i}}(\langle c_{k+Q\uparrow }^{\dagger }c_{-k\downarrow
}^{\dagger }\rangle  -\langle c_{k\uparrow }^{\dagger }c_{-(k+Q)\downarrow }^{\dagger }\rangle
)] \nonumber \\
&=&\varphi^{0} + e^{i
{\bf Q\cdot R}_{i}}\varphi^{Q},  \label{5}
\end{eqnarray}
where $\sum_{k}^{\prime }$ denotes a summation over $k$ within the reduced
Brillouin zone. $\varphi^{Q}$ is, in general, a complex quantity. There
is also an imaginary term in $\tau $, which involves mean values of the form 
$\langle c_{k+Q\sigma }^{\dagger }c_{k\sigma }\rangle $. However, all the
parameters are real for the lowest-energy solution. 
Note that $\varphi^{Q}$ is the mean value of the $%
S_{z}=0$ projection of a nearest-neighbor triplet pair \cite{so5}. This is
a consequence of the breaking of time-reversal symmetry in the SDW state.

As usual, the normal terms of the mean-field Hamiltonian are diagonalized by
the canonical transformation: 
\begin{eqnarray}
c_{k\sigma } &=&u_{k}\gamma _{k\sigma }^{(-)}-\sigma v_{k}\gamma _{k\sigma
}^{(+)},\nonumber \\
u_{k}\;=\;\sqrt{\frac{1}{2}(1+\frac{\epsilon _{k}}{%
E_{k}})},  \nonumber \\
c_{k+Q\sigma } &=&\sigma v_{k}\gamma _{k\sigma }^{(-)}+\sigma u_{k}\gamma
_{k\sigma }^{(+)},\nonumber \\
v_{k}\;=\;\sqrt{\frac{1}{2}(1-\frac{\epsilon
_{k}}{E_{k}})},  \label{6}
\end{eqnarray}
where $\gamma _{k\sigma }^{(-)},\;\gamma _{k\sigma }^{(+)}$ are defined on
the valence and conduction bands with dispersion relations $E_{k}^{\pm }=\pm
E_{k}$, respectively, where $E_{k}=\sqrt{\epsilon _{k}^{2}+\Delta ^{2}}$, $%
\epsilon _{k}=-2t_{ef}(\cos k_{x}+\cos k_{y})$, and $\Delta
=m(U/2+4t_{3}\tau )$. The anomalous terms are not exactly diagonalized by (%
\ref{6}). For example 
\begin{eqnarray}
c_{k\uparrow }^{\dagger }c_{-k\downarrow }^{\dagger } &=& u_{k}^{2}\gamma
_{k\uparrow }^{(-)\dagger }\gamma _{-k\downarrow }^{(-)\dagger
}-v_{k}^{2}\gamma _{k\uparrow }^{(+)\dagger }\gamma _{-k\downarrow
}^{(+)\dagger }\nonumber\\
& +& u_{k}v_{k}(\gamma _{k\uparrow }^{(-)\dagger }\gamma
_{-k\downarrow }^{(+)\dagger }-\gamma _{k\uparrow }^{(+)\dagger }\gamma
_{-k\downarrow }^{(-)\dagger }).  \label{7}
\end{eqnarray}
Below half filling and for sufficiently large magnitude of the charge gap $%
\Delta $, the interband pairs can be neglected. In the two-band basis (\ref
{6}) the mean-field Hamiltonian (\ref{3}) results: 
\begin{equation}
H^{MF}\;=\;\sum_{k\alpha }\;[ \sum_{\sigma}(\xi _{k}^{\alpha }\gamma _{k\sigma
}^{\alpha \dagger }\gamma _{k\sigma }^{\alpha })-
(\Delta _{k}^{\alpha
s}\gamma _{k \uparrow}^{\alpha \dagger }\gamma _{-k \downarrow }^{\alpha \dagger
}+h.c.)],  \label{8}
\end{equation}
with $\alpha =+,-$, and 
\begin{eqnarray}
\xi _{k}^{\alpha }&=& E_{k}^{\alpha }-\mu _{ef},\nonumber\\
\Delta_{k}^{\alpha s}&=&(\alpha \;\varphi^{0}+2u_{k}v_{k}
\varphi^{Q})\;4t_{3}\tau (\cos k_{x}-\cos k_{y}).  \label{9}
\end{eqnarray}
After a Bogoliubov transformation, the set of self-consistent equations can
be cast as: 
\begin{eqnarray}
n &=&1-\frac{1}{L}\sum_{k\alpha }^{\prime }\;\frac{\xi _{k}^{\alpha }}{%
\lambda _{k}^{\alpha }}(1-2f(\lambda _{k}^{\alpha })),  \nonumber \\
\tau &=&\frac{1}{2L}\sum_{k\alpha }^{\prime }\;\frac{\epsilon _{k}\cos k_{x}%
}{E_{k}}\;\alpha \;\frac{\xi _{k}^{\alpha }}{\lambda _{k}^{\alpha }}%
(1-2f(\lambda _{k}^{\alpha })),  \nonumber \\
m &=&m\;(\frac{U}{2}+4t_{3}\tau )\;\frac{1}{2L}\;\sum_{k\alpha }^{\prime }\;%
\frac{1}{E_{k}}\;\alpha \;\frac{\xi _{k}^{\alpha }}{\lambda _{k}^{\alpha }}%
\;(1-2f(\lambda _{k}^{\alpha })),  \nonumber \\
\varphi^{0} &=&\frac{1}{2L}\sum_{k\alpha }^{\prime }\;\;\alpha
\;(1-2f(\lambda _{k}^{\alpha }))\;\frac{\Delta _{k}^{\alpha s}}{\lambda
_{k}^{\alpha }}\;\cos k_{x},  \nonumber \\
\varphi^{Q} &=&-\frac{1}{2L}\sum_{k\alpha }^{\prime
}\;2u_{k}v_{k}\;(1-2f(\lambda _{k}^{\alpha }))\;\frac{\Delta _{k}^{\alpha s}%
}{\lambda _{k}^{\alpha }}\;\cos k_{x},  \label{10}
\end{eqnarray}
where $\lambda _{k}^{\alpha }\;=\;\sqrt{(\xi _{k}^{\alpha })^{2}+(\Delta
_{k}^{\alpha s})^{2}}$ and $f(\lambda _{k}^{\alpha })\;=\;1/(1+\exp (\beta
\lambda _{k}^{\alpha }))$, with the temperature $T=1/\beta $ in units where
the Boltzmann constant $k_{B}=1$. Notice that in a SDW background 
$\varphi^{0},
\;\varphi^Q$ depend on $U$ through $u_{k},v_{k}$ and $%
\lambda _{k}^{\alpha }$. It is apparent from Eqs. (\ref{9}-\ref{10}) that
these quantities contribute with opposite signs to the superconducting gap.
At first sight it seems to be valid to neglect the contributions of the
triplets $\varphi^{Q}$ \cite{in} to the BCS gap. Under this assumption
the dependence of the mean-field critical temperature as a function of
doping is shown by the solid circles in Fig. 1. The maximum of $T_{c}$ occurs for values of the
chemical potential $\mu _{ef}$ equal to the energy of the van Hove
singularity of the density of states of the SDW solution. As this density of
states is enhanced with respect to the non-interacting one, the value of the
maximum $T_{c}$ is, thus, enhanced with respect to the value corresponding
to the paramagnetic solution for the same parameters. However, it can be
easily seen from Eq. (\ref{10}) that 
$\varphi^{0}\rightarrow -\varphi^{Q}$ as 
$U\gg t$. In other words, as the staggered magnetization
increases to values closer to that of the Neel state ($m=1$), the absolute
value of the contributions of triplet and singlet states become equal and
the superconducting solution disappears. In the weak coupling regime, the
superconducting gap does not vanish. It is, however, highly reduced with
respect to that of the paramagnetic phase. 
For example: for optimum doping, 
$\varphi^0 = 0.226 \times 10^{-2} ,\;\varphi^Q = -0.198
\times 10^{-2} $ for $U=6 t$,
 while $\varphi^0 = 0.122 \times 10^{-2},\;$
$\varphi^Q =-0.112 \times 10^{-2}$ for $U=8 t$.
When the contribution of triplets is neglected, 
$\varphi^0=0.283 \times 10^{-1}$
for optimum doping and $U=8t$. The maximum value of  $\varphi_x$
for the  paramagnetic solution is 
 $\varphi_x = 0.187 \times 10^{-1}$, independently of
the magnitude of $U$.
For $U > 8t$ we were not able to find the numerical solution of
$\varphi^0, \; \varphi^Q$.
One might expect that including
magnetic fluctuations, the singlet contribution, and with it the
superconducting order parameter is enhanced. This is discussed in the next
section.

\section{RPA fluctuations.}

The complete treatment of the fluctuations of the antiferromagnetic
background of the doped system is a quite hard task. The doping on the Neel
state is expected to introduce instabilities towards the formation of
incommensurate and spiral SDW kind of ordering \cite{SS} as well as quantum
transitions to a quantum-disordered phase with antiferromagnetic short-range
order, depending on the topology of the Fermi surface \cite{sach}. In this
section, we de not consider these effects at all, and we concentrate in the
study of the usual magnon SDWF, assuming that the doping does not affect the
long-range order of the background. Even when this assumption is only
strictly valid at half filling or for a strongly underdoped system, it is
frequently used in several theories of the high-$T_{c}$, as mentioned in the
introduction. For the Hubbard model at half filling, these SDWF within the
RPA approximation were shown to be enough to recover the value of the local
magnetization $\langle S_{z}\rangle $ of the Heisenberg model in the limit
of $U\gg t$ \cite{schri,chu-fre}. These fluctuations were further proposed
to mediate the pairing interaction in the SDW background \cite{schri,tej}.
However, a more detailed study of the effective interactions for the Hubbard
model \cite{fre,chu-fre} indicated that in fact they do not provide any
pairing mechanism. In what follows, we analyze if they can correct the
cancelation effect between pairs with total momentum $0$ and $Q$ presented
in the previous section. In the following, we restrict our study to $T=0$.

The three-body terms are reduced to two-body ones in this level of
approximation following a similar procedure as with the Hartree-Fock
decoupling \cite{fog}. It is found (up to a constant): 
\begin{eqnarray}
t_{3}\;\sum_{<ij>\sigma } &&[\frac{n}{2}\;c_{i\sigma }^{\dagger }c_{j\sigma
}\;(n_{i-\sigma }+n_{j-\sigma })\nonumber \\
& - & \tau \;(c_{j\sigma }^{\dagger }c_{i\sigma
}c_{j-\sigma }^{\dagger }c_{i-\sigma }\;+\;h.c)  \nonumber \\
+2 &&\tau \;(\;n_{i\sigma }n_{j\sigma }\;-\;c_{i\sigma }^{\dagger
}c_{j-\sigma }^{\dagger }c_{i-\sigma }c_{j\sigma })].  \label{11}
\end{eqnarray}
The terms of the first line of Eq. (\ref{11}) contribute only to the renormalization of the
band-width and of the chemical potential, and for low-densities to the BCS-$%
s $-wave solution, but they do not couple with the SDWF. In the SDW
background, the first term of the second line of Eq. (\ref{11}) gives the same
contribution as $4t_{3}\tau \;\sum_{<ij>}\;S_{i}^{z}S_{j}^{z}$ plus a
nearest-neighbor repulsion, which is also reflected in the 
expression of the
charge gap of the SDW solution. The second term, can be written as 
\begin{equation}
2 t_{3}\tau \;\sum_{<ij>}\;(S_{i}^{+}S_{j}^{-}+S_{i}^{-}S_{j}^{+}),
\label{12}
\end{equation}
with $S_{i}^{+}=c_{i\uparrow }^{\dagger }c_{i\downarrow }$. Putting both
latter contributions together, a Heisenberg interaction with
antiferromagnetic coupling $J=4t_{3}\tau $ is obtained. The pairing terms
come from Eq. (\ref{12}) and they couple with the transverse SDWF. Longitudinal
SDWF do not contribute to the pairing mechanism within the present approach.
Thus, we concentrate on the transverse channel. Writing Eq. (\ref{12}) in the
two-band SDW basis (Eq. (\ref{6})) and keeping only the terms of the valence band,
which are the relevant ones for the superconductivity bellow half filling,
it is found 
\begin{eqnarray}
& &\frac{1}{N}\;\sum_{k,q}^{\prime
}\;U_{q}\;(1\;-\;4u_{k}v_{k}u_{k+q}v_{k+q} )\nonumber\\
& &\gamma _{k+q\uparrow
}^{(-)\dagger }\gamma _{-(k+q)\downarrow }^{(-)\dagger }\gamma
_{-k\downarrow }^{(-)}\gamma _{k\uparrow }^{(-)},  \label{13}
\end{eqnarray}
with 
\begin{equation}
U_{q}=-4t_{3}\tau \;(\cos q_{x}+\cos q_{y}).  \label{bare}
\end{equation}
While the first term contains the contributions of pairs with $0$-momentum: 
\begin{eqnarray}
& & c_{k\uparrow }^{\dagger }c_{-k\downarrow }^{\dagger }c_{-(k+q)\downarrow
}c_{k+q\uparrow },\nonumber\\
& & c_{k+Q\uparrow }^{\dagger }c_{-(k+Q)\downarrow
}^{\dagger }c_{-(k+q+Q)\downarrow }c_{(k+q+Q)\uparrow },
\end{eqnarray}
the second term is obtained from the contributions of pairs of triplets with
momentum $Q$: 
\begin{equation}
c_{k\uparrow }^{\dagger }c_{-(k+Q)\downarrow }^{\dagger }c_{-(k+q+Q)\downarrow
}c_{(k+q)\uparrow }.
\end{equation}
As discussed in the previous section, both terms tend to cancel each other.

The spin susceptibility of the SDW state is a $2\times 2$ matrix with
diagonal elements $\chi ^{+-}(q,q;\omega ),\;\chi ^{+-}(q+Q,q+Q;\omega )$
and off-diagonal ones $\chi ^{+-}(q+Q,q;\omega ),\;\chi ^{+-}(q,q+Q;\omega )$
\cite{schri,tej,fre,chu-fre}. The matrix elements of the bare susceptibility
with total momentum ${\bf Q}$ (off-diagonal matrix elements) do not
contribute in the static case considered here \cite{sach}. Away from
half filling, the full RPA susceptibility, has interband as well as
intraband contributions. The Goldstone modes of the Neel state are obtained
from the poles of the interband part of the full susceptibility. In this
case from the solution of: 
\begin{equation}
1-(U-4t_{3}\tau U_{q})\chi _{0}^{+-}(q,\omega )=0,\;\;\;\;\;\;\omega \sim
0,\;\;\;\;\;\;{\bf q}\sim {\bf Q},  \label{14}
\end{equation}
with $U_{q}$ given in Eq. (\ref{bare}) and $\chi _{0}^{+-}(q,\omega )$, the
diagonal matrix elements of the bare susceptibility. This equation reduces
to the gap equation of the SDW mean-field solution, for $\omega =0,\;{\bf q}=%
{\bf Q}$, indicating the consistency of the approach. The intraband
contribution accounts for the instabilities of the SDW antiferromagnetic
background with long-range order and for the transition to the
quantum-disordered state \cite{sach,chu-fre} and we do not consider them in
the present work. When treated within the magnon-pole approximation \cite
{schri,chu-fre}, the interband terms result 
\begin{eqnarray}
\chi ^{+-}(q,q;\omega ) &=&-\frac{1}{2}\;\sqrt{\frac{1-\eta _{q}}{1+\eta _{q}%
}}\;[\frac{1}{\omega -\Omega _{q}+i\delta }\nonumber\\
& - &\frac{1}{\omega +\Omega
_{q}-i\delta }]  \nonumber \\
\chi ^{+-}(q,q+Q;\omega ) &=&-\frac{1}{2}\;[\frac{1}{\omega -\Omega
_{q}+i\delta }\nonumber \\
&+&\frac{1}{\omega +\Omega _{q}-i\delta }],  \label{15}
\end{eqnarray}
with $\Omega _{q}=2J_{ef}\;\sqrt{1-\eta _{q}^{2}}$, and $\eta _{q}=(\cos
q_{x}+\cos q_{y})/2$. The effective antiferromagnetic exchange coupling is
defined from the dispersion relation of the spin waves. Following usual
procedures \cite{schri,tej,chu-fre}, it is obtained for the present case $%
J_{ef}=4t_{ef}^{2}/U+4t_{3}\tau $. From Eq. (\ref{10}), it can be seen that at
half filling $\tau \sim \langle cosk_{x}\rangle /\Delta $ $\sim 2/(\pi
^{2}\Delta )$. Thus, in the limit $U\rightarrow \infty $, $t_{ef}\rightarrow
t_{AB}$ and $J_{ef}\rightarrow 4t_{AB}^{2}/U$, as expected from the results
of a canonical transformation on Eq. (1) for large $U$.

The fermion-fermion interactions constructed from the corresponding
fermion-magnon ones (see Fig. 8 of Ref. \cite{schri}) are: 
\begin{eqnarray}
&&\frac{1}{N}\sum_{k,q}^{\prime
}\;[-\;f_{1}(k,q)\;V(q,q)\;+\;f_{2}(k,q)\;V(q+Q,q+Q)  \nonumber \\
&&+\;f_{3}(k,q)\;(V(q,q)-V(q+Q,q+Q))] \nonumber\\
& & \gamma _{k+q\uparrow }^{(-)\dagger
}\gamma _{-(k+q)\downarrow }^{(-)\dagger }\gamma _{-k\downarrow
}^{(-)}\gamma _{k\uparrow }^{(-)},  \label{16}
\end{eqnarray}
where 
\begin{eqnarray}
& &f_{1}(k,q)\;=\;u_{k+q}^{2}u_{k}^{2}+v_{k+q}^{2}v_{k}^{2},\nonumber \\
& &f_{2}(k,q)\;=\;v_{k+q}^{2}u_{k}^{2}+u_{k+q}^{2}v_{k}^{2},  \nonumber \\
& &f_{3}(k,q)\;=\;2\;u_{k+q}v_{k+q}u_{k}v_{k}, \nonumber\\
& &V(q,q)=(U-U_{q})^{2}\chi ^{+-}(q,q;\omega =0).
\end{eqnarray}
As in the case of the bare interaction, we kept only terms on the valence
band. The first (second) line of Eq. (\ref{16}) contains the contributions of
pairs with total momentum $0$ ($Q$). The bare interaction (\ref{bare}) has
components in the $d$-wave- as well as in the $s $-wave-channel. The
relevant terms for the $d$-wave part, are those with ${\bf k}$ near $(0,\pi
) $ and ${\bf q}$ near $(0,0)$ as well as symmetry-related points. SDWF
would help to superconductivity in the case that the second line of Eq. (\ref{16}%
) tends to cancel the second term of Eq. (\ref{13}) while the first line of Eq.  (\ref{16}) vanishes or has the same sign 
as the first term of Eq. (\ref{13}) for
these wave vectors. For $\Delta \gg t$, $E_{k}\sim \Delta $ and $%
f_{3}(k,q)\rightarrow 1/2$. Expanding the structure factors $%
f_{1}(k,q),\;f_{2}(k,q)$, the first line of Eq. (\ref{16}) can be written as 
\begin{eqnarray}
& &-\frac{1}{2}\;(V(q,q)-V(q+Q,q+Q))\;- \nonumber\\
& &\frac{\epsilon _{k}\epsilon _{k+q}}{
\Delta ^{2}}\;(V(q,q)+V(q+Q,q+Q)).  \label{agre}
\end{eqnarray}
The first term of Eq. (\ref{agre}) exactly cancels the second line of Eq. (\ref{16}).
 The remaining term, which comes from pairs with momentum $0$, can be
expanded for small ${\bf q}$, 
\begin{eqnarray}
V(q,q)+V(q+Q,q+Q) &\sim &\frac{2}{J_{ef}}\;[\;\frac{(U+8t_{3}\tau )^{2}}{%
q_{x}^{2}+q_{y}^{2}} \nonumber\\
&-&8t_{3}\tau (4t_{3}\tau +U)\;]  \nonumber \\
\frac{\epsilon _{k}\epsilon _{k^{\prime }}}{\Delta ^{2}} &\sim &\frac{({\bf %
v_{k}}\cdot \delta {\bf k})^{2}}{\Delta ^{2}},
\end{eqnarray}
where 
$v_{k}^{i}=\partial \epsilon _{k}/\partial k_{i}|_{{\bf k}_{0}}$
and $\delta {\bf k}\sim \delta {\bf k}^{\prime }={\bf k}-{\bf k}_{0} \sim 
{\bf q}$. The latter quantity is vanishingly small except for ${\bf k}_0
\sim (\pi /2,\pi /2)$, in which case \cite{fre,chu-fre}, the effective
interaction in real space is a local repulsion in the triplet-channel plus a
long-range dipolar interaction and it is expected to induce spiral
distortions in the antiferromagnetic state \cite{SS}.

Thus, SDWF of the Neel state treated in the RPA approximation, do not modify
the picture obtained at the mean-field level.

\section{Conclusions and discussion.}

Our results can be summarized as follows. We studied an effective one-band
model for the superconducting cuprates in the underdoped regime. The model
has an effective pairing interaction and a BCS- mean-field solution with $d$%
-wave symmetry in the range of doping $0<\delta <.5$, which might be
relevant for the pairing mechanism of the cuprates. However, in the
mean-field level, this solution has always higher energy than the
antiferromagnetic SDW one. For the case of the $t-J$ model, for which the
results of strong-coupling mean-field \cite{t-J1} as well as numerical
techniques \cite{t-J2}, indicate that it exhibits superconductivity, the
instability of the Neel state is found also at very high doping within some
mean field approaches \cite{dop}. In other more phenomenological approaches,
antiferromagnetic long-range order has been used to simulate the
short range antiferromagnetic correlations \cite{naz,fei}. In this work we
explored the possibility of coexistence of both kinds of order:
antiferromagnetism and superconductivity. We found that neglecting the
contribution of triplet-pairs with momentum $Q$ as done in other approaches
for simplicity \cite{in}, there is an enhancement of the $d$-wave
superconducting solution due to the modified density of states of the
antiferromagnetic state with long-range order. We found, however, that both
contributions are equally important and tend to cancel each other. For large
values of the Coulomb repulsion $U$ there is no chance for superconductivity
and for low to intermediate values the magnitude of the superconducting gap
is much weaker than that of the paramagnetic BCS solution.

Our results are more transparent when analyzed in real space. The terms of
the Hamiltonian here considered, that cause the pairing, are the three-body
ones. Treating them in mean-field and reducing them to two-body ones, a
nearest-neighbor spin-flip antiferromagnetic interaction is found. This kind
of interaction is expected to generate an effective attraction for
nearest-neighbor spins in the singlet channel. In fact, in the
strong-coupling limit, resonating valence bond (RVB) singlets, which are
widely accepted to build up a superconducting state, are mainly a
consequence of the same kind of interaction \cite{an,rvb}. The formation of
singlet pairs is, however, very unlikely in an antiferromagnetic background
with long-range order. Spin fluctuations of the Neel state in the RPA level
(interband fluctuations) do not correct this effect. For higher doping,
intraband fluctuations are important and the antiferromagnetic correlations
are short-range like in real materials \cite{sach}. This scenario is better
described by a nearly antiferromagnetic Fermi liquid picture \cite{pines}.
We expect the pairing mechanism contained in this model to be active within
such a background. This is confirmed by preliminary results \cite{prog}.

In other theories, like the antiferromagnetic van Hove mechanism \cite
{naz,fei} or the spin-polaron model \cite{pla}, the superconductivity is
mainly determined by kinetic reasons. These theories assume that holes like
to propagate in the long-range antiferromagnetic background distorting it as
less as possible. In the first case, holes are assumed to move within the
same sublattice and to experience an attractive interaction $\sim
-0.6J=J\langle {\bf S}_{i}\cdot {\bf S}_{j}\rangle $ between
nearest-neighbor holes. Note however, that the results of the previous
section and previous works \cite{chu-fre} suggest that it is not valid to
replace the exchange interaction by a nearest-neighbor attraction in the
Neel state. Numerical calculations also show that the antiferromagnetic van
Hove scenario does not represent correctly the physics of realistic $t-J$%
-like models \cite{com}. The $t-J$ model contains, however, an explicit
nearest-neighbor attraction of magnitude $J/4$, which plays a relevant role
in most of the strong-coupling approaches \cite{an,rvb,t-J1,t-J2}. In the
case of the spin-polaron model \cite{pla}, this nearest-neighbor attraction
is neglected and hopping of a hole between both sublattices is allowed by
the occurrence of a local spin-flip in the slave-fermion representation
used. The holes become paired through the hole-magnon interaction originated
in the hopping process and the resulting $T_{c}$ in this strong-coupling
approach becomes appreciable in contrast to our weak coupling results and
those of Ref. \cite{chu-fre}. It is quite easy to inspect the correlated
hopping terms and observe that in the present model no terms like $%
c_{j\uparrow }^{\dagger }c_{i\uparrow }S_{i}^{+}$ are generated within the
mean-field treatment. Such terms could lead to a polaron-like picture in the
weak-coupling formalism.

A BCS mean-field approximation does not lead to superconductivity in
the ordinary Hubbard model ($t_{ab}=t$) with $U>0$.
There are instead, several calculations of the superconducting
gap with other mean-field Eliashberg-like theories,
based on the fluctuation-exchange approximation (FLEX)
\cite{flex}, which suggest $d$-wave superconductivity in the model.
However many
numerical attempts to find evidences of superconductivity
in the pure Hubbard model have been negative \cite{hubno}.
There are two features of the results based on the FLEX approximation
 that could lead one to speculate about some connection between these
treatments for the Hubbard model 
 and the results we presented here: a) within the FLEX approximation,
pairing is caused by an effective attraction in the $d$-wave
channel, which is
originated by spin fluctuations in the paramagnetic phase. 
In the present
case, pairing is originated by the interaction (\ref{12}), which is
precisely the nearest-neighbors 
static version of an effective attraction
mediated by the transverse spin fluctuations. 
Thus, within the paramagnetic phase,
superconductivity in the present model could have a similar origin as that
of FLEX for the pure Hubbard model.
b) As discussed in detail in Ref. \cite{viltre}, the FLEX approximation offers
a quite poor treatment of antiferromagnetic correlations. In particular,
the antiferromagnetic state is not recovered, even at half filling.
The behavior of the superconducting gap within the FLEX approximation
\cite{flex}
is very similar to that shown in Fig. 1 for the BCS $d$-wave solution
for $t_{AB} \neq 0$ in the paramagnetic phase.  
Questions arise about what could occur in the pure Hubbard model,
in the case that the tendency towards long-range antiferromagnetism 
could be also included in some FLEX-like scheme. 
In connection with this latter point, it could be stressed that
the effective interactions
generated by the SDWF of the antiferromagnetic
state of the Hubbard model \cite{schri,tej,fre,chu-fre},
 where shown not to be able to provide
any effective attractive interaction in the $d$-wave channel 
\cite{fre,chu-fre}.

\section*{Acknowledgments.}
 We thank useful discussions with Prof. Fulde and G.
Zwicknagl. L. A. thanks A. Ruckenstein for bringing Ref.\cite{in} to our
attention. During most of this work L.A. was an Alexander von Humboldt
fellow. A.A.A. is partially supported by the Consejo Nacional de
Investigaciones Cient\'{\i }ficas y T\'{e}cnicas (CONICET), Argentina. We
thank Prof. Bennemann for his hospitality.


{\large {\bf FIGURE CAPTIONS}}

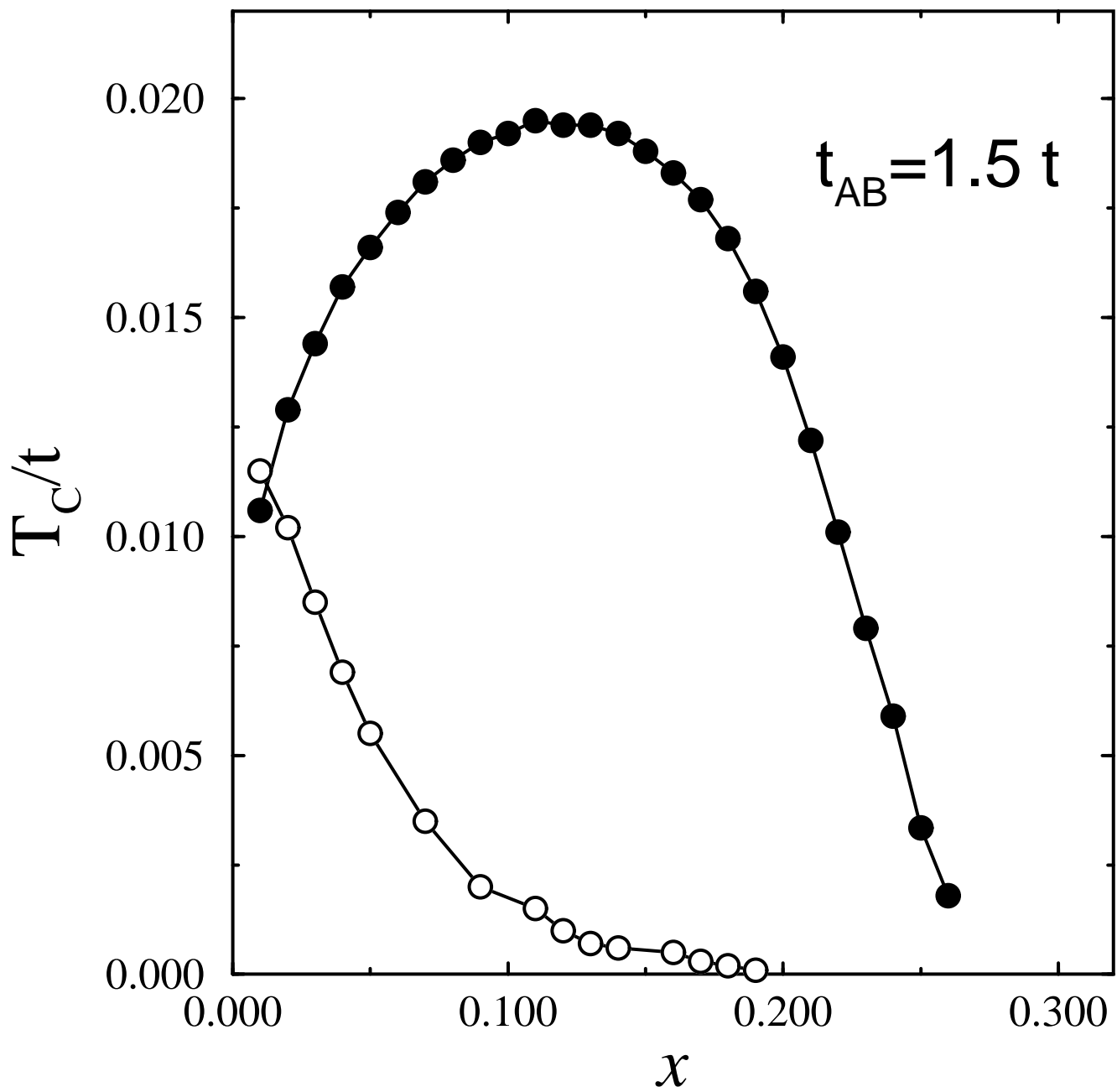
{\bf Fig. 1.} Mean-field critical temperature as a function of doping $x=1-n$
for $t_{AA}=t_{BB}=t$ and $t_{AB}=1.5t$.  Open circles correspond to the
usual paramagnetic $d$-wave BCS solution. Solid circles correspond to the $d$%
-wave BCS solution in the SDW background for $U=8 t$, neglecting the
contributions of triplets with total momentum ${\bf Q}=(\pi ,\pi )$.

\end{document}